\documentclass[final,authoryear,5p]{elsarticle}

\usepackage{graphicx}
\usepackage{multicol}

\usepackage{amssymb}
\usepackage[fleqn]{amsmath}

\usepackage[ps2pdf,%
a4paper=true,%
breaklinks=true,%
colorlinks=true,%
pdfauthor={First Author et al.},%
pdftitle={Template for manuscripts in Advances in Space Research}%
]{hyperref}

\journal{Advances in Space Research}

\begin{document}

\begin{frontmatter}



\title{Cosmic-ray diffusion in collisionless plasmas including pressure anisotropy}


\author{M. S. Nakwacki\corref{cor}}
\address{Instituto de Astronom\'ia y F\'isica del Espacio (IAFE--CONICET), Cuidad Universitaria, Buenos Aires 1428, Argentina}
\cortext[cor]{Corresponding author}
\ead{sole@iafe.uba.ar}



\author{J. Peralta-Ramos}
\address{Departamento de F\'isica, Facultad de Ciencias Exactas y Naturales, Universidad de Buenos Aires and IFIBA--CONICET, Cuidad Universitaria, Buenos Aires 1428, Argentina}
\ead{jperalta@df.uba.ar}

\begin{abstract}

Using a hybrid kinetic magnetohydrodynamic formalism incorporating the effects of pressure anisotropy, we simulate the evolution of a turbulent collisionless plasma in six different models covering the sub/super-sonic and sub/super-Alfv\'enic regimes. Based on the power spectrum of the simulated magnetic field, we compute the particle diffusion coefficients for protons with kinetic energy in the $50-500$ MeV range, and compare them to those obtained within standard magnetohydrodynamics. Our results show that the differences in the  statistical properties of the magnetic field, generated by pressure anisotropy and its associated kinetic instabilities, have an appreciable impact on the diffusion coefficients of energetic protons. Moreover, the values of the diffusion coefficients that we obtain within each of the six models considered vary significantly.

\end{abstract}


\end{frontmatter}

\parindent=0.5 cm


\section{Introduction}

A complete understanding of the transport of charged energetic particles through a magnetized turbulent plasma poses 
a challenge in many areas, ranging from fusion research to space physics. The main difficulty arises due to the stochastic nature of the magnetic field. 
In particular, the diffusion of charged particles in a fluctuating magnetic field has attracted a lot of attention due to its relation with cosmic ray transport.
Different approaches have been developed to study particle diffusion in turbulent plasmas, among which the most well-known are the quasilinear theory (QLT) 
\citep{Jokipii66,Shalchi05b,Dosch13,Schlick1998}, hard-sphere scattering models \citep{Gleeson69}, nonlinear guiding center formalisms \citep{Matt2003,Shalchi08}, approaches based on velocity correlation functions \citep{Bieber97}, and test-particle simulations \citep{Giacalone99,Gao11,Xu2013,Tautz2013a,Tautz2013b,Casse2012,Guo2010}. For reviews on energetic particle transport through space plasmas see e.g. \citet{Jokipii71,Giacalone98,JoeBOOK,Parker65,Zweibel2013,Jokipii2008,Jokipii2010,Schlick94,Potgieter13,Carbone2013} and references therein.

The diffusion of charged particles in a turbulent plasma is directly related to the statistical properties of the magnetic field. In weakly collisional plasmas, particles can exhibit anisotropic distributions
with respect to the local magnetic field direction, that can survive for considerably long periods compared to dynamical timescales of certain systems. The interplay between temperature anisotropy, the kinetic instabilities that are triggered by it, and strong turbulence leads to complex phenomena in weakly collisional space plasmas  that can affect their macroscopic evolution in nontrivial ways \citep{Lazarian12,Bran2013,Matthaeus95,Matteini07,Maruca11,Bale09,Chen10,Samsonov07,Osman13,
Matteini06,Osman12,Servidio13,Hellinger2008,Schekochihin10,Kunz11}.

Collisionless plasmas are often described by the Maxwell-Vlasov equation, which is computationally quite demanding to solve. Fortunately, many properties of weakly collisional plasmas can be reasonably well described by fluid-like models, provided that some additional constrains are considered. The simplest fluid-like approximation is the well-known double adiabatic MHD model developed by Chew, Goldberg and Low (CGL)
\citep{Chew56}. A modified CGL-MHD model incorporating the anisotropy constraints due to 
kinetic instabilities has been used for modeling the solar wind and magnetosphere in numerical simulations
\citep{Samsonov01,Samsonov07,Meng12a,Meng12b,dza1}. Similar models have been used in \citet{Kunz11} to study the heating of central regions
of cold-core clusters of galaxies, in \citet{Sharma07} to investigate 
the magnetorotational instability in accretion disks
around black holes, and in \citet{SantosLima2011} to simulate a turbulent dynamo. In \citet{Kowal11}, 
it was shown that the presence of the instabilities driven by temperature anisotropy has strong impact on the evolution of the plasma density, velocity and magnetic field.
More recently, CGL-MHD numerical simulations were performed \citep{Nakwacki12,SantosLima2013} to study fast growing magnetic fluctuations in the smallest scales which operate in the collisionless plasma that fills the intra-cluster medium. A comparison of Faraday Rotation Maps obtained from CGL-MHD and MHD simulations was carried out in Ref. \citet{faraday}. 

It turns out that the statistical properties of the turbulent magnetic field obtained from the CGL-MHD formalism are in general quite different from those obtained from MHD (except in the high plasma $\beta$ plasma case, see \citet{SantosLima2013}). These differences are largely due to the effect of kinetic instabilities such as the firehose and mirror instabilities, both of them triggered by the pressure anisotropy present in collisionless plasmas.
It is therefore of interest to quantify the impact of differences in the statistical properties of the magnetic field (that result from the presence of pressure anisotropy) on the transport of energetic particles through a magnetized fluctuating plasma. 

In this paper, we perfom CGL-MHD and MHD 3D numerical simulations of the turbulent magnetic field, using different plasma parameters that cover the sub/super-Alfv\'enic and the sub/supersonic regimes. Using the simulated magnetic field, we calculate the diffusion coeficients for protons with kinetic energies in the $50-500$ MeV range. Since our aim here is to determine the impact of pressure anisotropy on the diffusion coefficients for particle propagation through a magnetized plasma in the simplest possible setting, we shall stick to the QLT. We must point out that in recent years the QLT has been shown to present several limitations (see, e.g., \citet{Dosch09,JoeBOOK,Bieber97,Tautz2013a,Tautz2013b}). 
The use of more elaborate and accurate formalisms to model particle transport in turbulent plasmas is left for future work. 

We should note that, to the best of our knowledge, the comparison of diffusion coeficients computed from CGL-MHD and MHD simulations has not been attempted before. In Ref. \citet{bete2011}, the CGL-MHD model was used to study particle acceleration in the intergalactic medium, which is nearly collisionless, taking into account kinetic effects that affect the turbulent magnetic field distribution by making it much more wrinkled than in a standard MHD turbulent system, and therefore substancially affecting cosmic-ray propagation. 

The paper is organized as follows. In Section \ref{theo} we briefly outline the QLT derivation of the diffusion coefficients for energetic particles in a fluctuating magnetic field, and also provide details of the numerical scheme used to simulate the evolution of the anisotropic plasma. In Section \ref{res} we present and discuss our results, and in Section \ref{con} we give our conclusions and outlook. 

\section{Theoretical setup}
\label{theo}

\subsection{Diffusion coefficients in quasilinear theory}
\label{dif}

The diffusion coefficients $D_\parallel$ and $D_\perp$ are computed from the first-order QLT for axisymmetric turbulence originally derived from the Fokker-Planck equation by \citet{Jokipii66,Jokipii71}. 
The validity of the QLT is discussed in detail in \citet{Jokipii66,Jokipii71} (see also \citet{Shalchi05b,Giacalone98}), but for clarity we will enumerate the assumptions made in its derivation. First, the fluctuating magnetic field is such that $\left\langle (\delta B)^2 \right\rangle/\left\langle B_0^2\right\rangle  \ll 1$, where $B_0$ and $\delta B$ are the mean magnetic field and a typical fluctuation of the magnetic field, respectively. This implies that particle motion can be well described by the two-point correlation of the magnetic field. Second, the changes in pitch angle of the particles due to scattering with magnetic irregularities are assumed to be small with respect to the correlation length of the magnetic field, and thus the pitch angle distribution can be taken as isotropic. And lastly, the variation of the particle distribution function $n$ along the direction of the particle's path is small over the distance in which a particle is scattered appreciably in pitch angle. The latter requirement implies that the time variation of $n$ is due to diffusion motion, which is much slower that individual particle motion.

For completeness, we shall briefly state the main results of the QLT that we will use throughout. We assume that there is a mean field $B_0(\vec{r}) \hat{z}$, and consider random fluctuations $\vec{B}_1(\vec{r}) = B_{1x}\hat{x}+B_{1y}\hat{y}+B_{1z}\hat{z}$, so that $\vec{B}(\vec{r})=B_0(\vec{r}) \hat{z} +\vec{B}_1(\vec{r})$. Define $\omega = \frac{Ze}{\gamma m_0 c} \vec{B}$, where $Ze$ and $m_0$ are the particle's charge and rest mass, $c$ is the speed of light, and $\gamma=(1-V^2/c^2)^{-1/2}$ with $V$ the particle's speed.
The basic quantity is the two-point correlation function
\begin{equation}
R_{ij}(\eta,\psi,\zeta)=\left\langle \omega_{1i}(x,y,z)\omega_{1j}(x+\eta,y+\psi,z+\zeta) \right\rangle
\end{equation}
In what follows, we will compute the diffusion coefficients along given trajectories, which for simplicity take to be parallel to the $\hat{z}$ axis. This corresponds in an approximate way to the situation we are interested in, namely one in which a spacecraft measuring the magnetic field travels through the plasma along these trajectories. 
To this end, we will need the one-dimensional power spectrum, which can be calculated in the plasma rest frame as follows (assuming the spacecraft travels at a constant speed $V_w$ in the $\hat{z}$ direction) 
\begin{equation}
P_{ij}(f)=\frac{\gamma^2 m_0^2 c^2}{Z^2 e^2}\int_{-\infty}^\infty R_{ij}(0,0,\zeta)e^{-2\pi i f\zeta /V_w} \frac{d\zeta}{V_w}
\label{pij}
\end{equation} 
The diffusion coefficients are then given by \citet{Jokipii71}.

\begin{equation}
D_{\perp} = \frac{1}{2}\int_0^1 \frac{\left\langle (\Delta x)^2 \right\rangle}{\Delta t} d\mu  
\end{equation}
and 
\begin{equation}
D_\parallel = \frac{2V^2}{9}\left[\int_0^1 \frac{\left\langle (\Delta \mu)^2 \right\rangle}{\Delta t} \right]^{-1}
\end{equation}
where 

\begin{equation}
\frac{\left\langle (\Delta \mu)^2 \right\rangle}{\Delta t} =\frac{Z^2 e^2}{\gamma^2m_0^2 c^2} \frac{(1-\mu^2)V_w}{|\mu|V}P_{XX}(f^*)
\end{equation}
and 
\begin{equation}
\frac{\left\langle (\Delta x)^2 \right\rangle}{\Delta t} = \frac{Z^2e^2}{\gamma^2 m_0^2 c^2} \frac{|\mu|V V_w}{2\omega_0^2}\left[2P_{XX}(0)+
\frac{1-\mu^2}{\mu^2}P_{ZZ}(f^*)\right]
\end{equation}
In these expressions, $\mu=V_z/V$ is the cosine of the particle's pitch angle and   
\begin{equation}
f^* = \frac{V_w \omega_0}{2\pi\mu V}
\end{equation}
is the resonant frequency.

\subsection{Numerical simulations of the turbulent magnetic field}
\label{num}

To simulate the dynamics of the collisionless magnetized plasma in the presence of pressure anisotropy, we use an MHD formalism with a Chew-Golberger-Low double-isothermal closure \citep{Chew56}, as implemented in the numerical code developed in \citet{Kowal11}. 

The equations for the magnetic field $\vec{B}$, particle velocity $\vec{V}$ and density $\rho$ can be written as

\begin{equation}
\frac{\partial \rho}{\partial t} + \vec{\nabla} \cdot (\rho \vec{V}) = 0
\end{equation}
\begin{equation}
\frac{\partial \vec{B}}{\partial t} - \vec{\nabla} \times (\vec{V} \times  \vec{B}) = 0
\end{equation}
and 
\begin{equation}
\frac{\partial (\rho \vec{V})}{\partial t} + \vec{\nabla} \cdot \left[ \left(a_\perp^2\rho + \frac{B^2}{8\pi} \right)\mathcal{I} - (1-\alpha)\vec{B}\vec{B}  \right] = \vec{f}
\end{equation}

The term $\vec{f}$ is a random solenoidal large-scale driving force representing the turbulence
driving, which is driven at wave scale $k = 2.5$ (i.e $2.5$ times smaller
than the size of the box used in the simulations). $P = p_\perp \mathcal{I} + (p_\parallel - p_\perp)\hat{b}\hat{b}$ 
is the pressure tensor with components $p_\parallel=a_\parallel^2 \rho$ and 
$p_\perp=a_\perp^2 \rho$ parallel and perpendicular to the magnetic field direction $\hat{b} = \vec{B}/|\vec{B}|$ ($\mathcal{I}$ stands for the unit matrix).  $a_\parallel$ and $a_\perp$ are constants and represent speed of sounds along the parallel and perpendicular directions to the magnetic field. 
The quantity $\alpha$ is defined as $\alpha = (p_\parallel - p_\perp)/(2P_{\textrm{mag}})$, where $P_{\textrm{mag}} = B^2/(8\pi)$ is the magnetic pressure.

We do not take into account viscosity and diffusion in the equations. The
numerical integration of the system evolution governed by the CGL-MHD equations were performed
by using the second-order shock-capturing Godunov-scheme code and the time integration was
done with the second-order Runge-Kutta method. Further details about the CGL-MHD code used here can be found in \citet{Kowal11} (see also \citet{SantosLima2011,SantosLima2013,Nakwacki12}). 

In order to study the dependence of the diffusion coefficients on the structure of the magnetic field, we have performed simulations for six different initial conditions covering the sub/super-Alfv\'enic and the sub/supersonic regimes, as indicated in Table \ref{table1}. The models can be further divided into strong or weak turbulence regimes, corresponding to $\rho \delta v^2 > p_\parallel \delta B^2$ and $\rho \delta v^2 < p_\parallel \delta B^2$, respectively, where $\delta v$ and $\delta B$ are root mean square values of velocity and magnetic field intensity. 
In all of the cases studied, the initial density is set equal to $5$ cm$^{-3}$, while the initial particle velocity is zero. In all of our numerical simulations we use a box of $0.1$ AU and a mesh-grid of $256^3$ points. 
The magnetic field is a sum of a mean field and fluctuations, i.e. $\vec{B}=\vec{B}_0 + \delta\vec{B}$. Initially, $\delta\vec{B}=0$. The mean magnetic field is set in the $\hat{z}$ direction, $\vec{B}_0 = B_0 \hat{z}$. The anisotropy parameter $\alpha$ is taken to be time-independent. We note that the values of the squared sound speed $c_s^2$ corresponding to each model is the same in the MHD and the CGL-MHD simulations.

\begin{table}[h]
\caption{Initial conditions and some features of the six models studied. The initial magnitude of the magnetic field $B_0$ is given in nT. supS/subS stand for supersonic/subsonic, while supA/subA stand for super-Alfv\'enic/sub-Alfv\'enic.}
\begin{tabular}{cccccc}
\hline
Model & $B_0$ & $a_\parallel$ & $a_\perp$ & Regime & Instability\\
\hline
1 & 5.0 & 1.0 & 2.0 & subS-subA & Mirror\\
2 & 5.0 & 1.0 & 0.5 & subS-subA & Firehose\\
3 & 0.5 & 0.1 & 0.2 & supS-supA & Mirror \\
4 & 0.5 & 0.1 & 0.05 & supS-supA & Firehose\\
5 & 0.5 & 1.0 & 0.5 & subS-supA & Firehose\\
6 & 5.0 & 0.1 & 0.2 & supS-subA & Mirror\\
\hline
\end{tabular}
\label{table1}
\end{table}

Models 1 and 2 correspond to weak turbulence in the subsonic and sub-Alfv\'enic regime. The difference between these two cases is that in model 1 mirror instabilities are triggered by pressure anisotropy, while in model 2 firehose instabilities are triggered. Models 3 and 4 correspond to strong turbulence in the supersonic and super-Alfv\'enic regimes, and differ with each other in the kind of kinetic instability that is present. Mirror and firehose instabilities arise in models 3 and 4, respectively. We note that models 3 and 4 correspond to the conditions prevailing in the solar wind. Finally, model 5 corresponds to the subsonic and super-Alfv\'enic regime and presents firehose instabilities in weak turbulence, while model 6 corresponds to the supersonic and sub-Alfv\'enic regime and presents mirror instabilities in strong turbulence.

To compute the diffusion coefficients, we use the magnetic field at a time corresponding to a fully developed turbulent cascade. We have averaged out the values of $D_\parallel$ and $D_\perp$ obtained from the $36$ trajectories closest to the center of the $x-y$ plane and parallel to the $\hat{z}$ axis. We have checked that reasonable convergence in these values is achieved provided that more than $\sim 16$ trajectories are considered in the average. For all of the models studied here, the standard deviation obtained for this set of $36$ trajectories is less than a tenth of the value of the diffusion coefficients, which sets our resolution for $D_\parallel$ and $D_\perp$.

\section{Results and discussion}
\label{res}

\subsection{Magnetic field}

In this section we shall briefly discuss the main differences between the plasma configuration attained at large times in the six models described in Table \ref{table1}. We recall that at this stage of the evolution of the plasma, the turbulent cascade is fully developed. 
We shall focus on the features of the magnetic field in each regime, since this is the physical quantity that is most directly connected to the diffusion coefficients, but to provide a more complete overview we will also comment on the particle's velocity and density distributions. Detailed discussions about the statistical properties of the magnetic field, the velocity and the density for these models in MHD and CGL-MHD can be found in \citet{Kowal11}. 

\begin{figure}[htpb]
\resizebox{\hsize}{!}{
\includegraphics{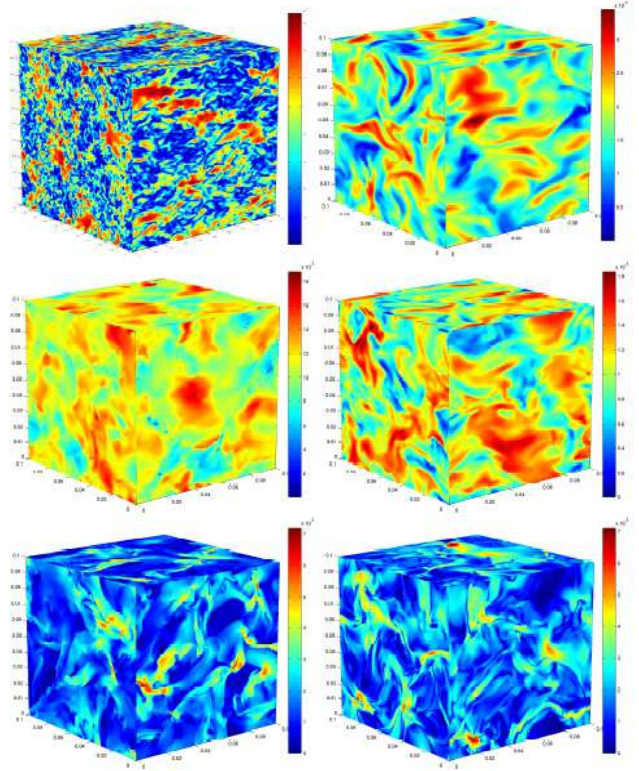}}
\caption{Central slices of computational domain of the magnetic field intensity (CGL-MHD left and MHD right)  for models 1, 2, and 3. }\label{f1b1}
\end{figure}

\begin{figure}[htpb]
\resizebox{\hsize}{!}{
\includegraphics{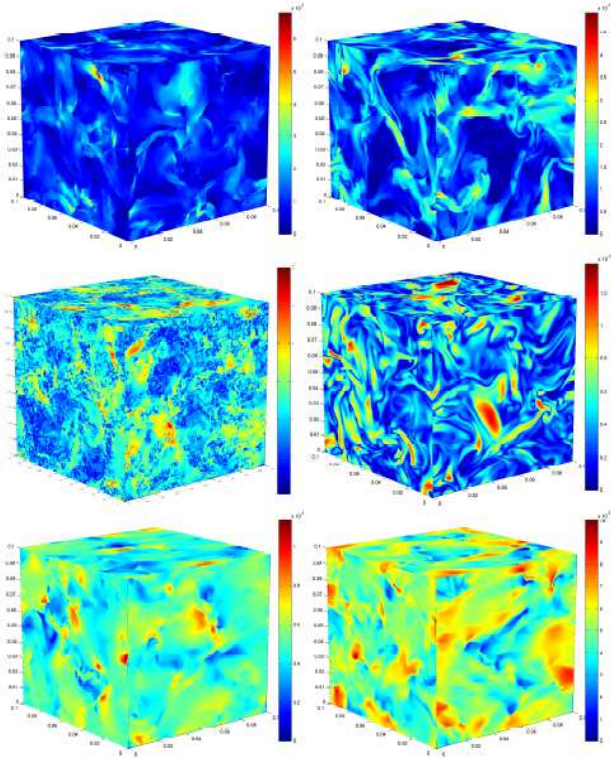}}
\caption{Central slices of computational domain of the magnetic
field intensity (CGL-MHD left and MHD right)  for models 4, 5, and 6. }\label{f1b2}
\end{figure}

Figures \ref{f1b1} and \ref{f1b2} depict the magnetic field intensity in the center of the 
computational domain for models 1, 2, and 3, and for models 4, 5, 6, respectively. The CGL-MHD results are shown in the left column, while the standard MHD case is shown in the right column. 

In the case of weak turbulence in the subsonic and sub-Alfv\'enic regime (first 
row of Figure \ref{f1b1}), corresponding to model 1, the presence of mirror instabilities produces changes in the velocity distribution, accelerating the plasma along the magnetic field lines resulting in the 
increase of the effective sonic Mach number and more elongated structures. As the turbulence is weak, the instabilities can grow without being suppressed by the turbulent motions of the gas. This results in large differences between the magnetic field structure obtained in CGL-MHD and in MHD.


The subsonic, sub-Alfv\'enic weak turbulence regime corresponding to model 2 is shown in the second row of Figure  \ref{f1b1}. This case presents 
strong firehose instabilities responsible for deformation of the magnetic field
lines. The curved magnetic lines tend to slow down and trap the flowing gas. Since the growth
rate is larger at small scales, we expect this effect to create more granulated maps. As in model 1, the instabilities can freely grow without being suppressed
by the turbulent motions of the gas. Here, the firehose
instability is responsible for the generation of small-scale magnetic field fluctuations that tangle
the field lines and result in an increase of the perpendicular pressure, thus leading to the isotropization of
the fluctuations with respect to the magnetic field lines. Comparing the MHD and CGL-MHD cases, it is seen that in the former the magnetic field intensity is more elongated. 

In the case of model 3 (supersonic and super-Alfv\'enic), shown in the third row of Figure \ref{f1b1}, the granulated structures are not present. The strong turbulence is able to destroy the magnetic field's configuration attained by the mirror instabilities.

As in the previous case, model 4, corresponding to the supersonic and super-Alfv\'enic regime with firehose instabilities, does not present granulated structure, as shown in the first row of Figure \ref{f1b2}. Instead, the magnetic field intensity presents a more homogeneous structure. Note that the structure of the magnetic field obtained in CGL-MHD for models 3 and 4 is very similar to the one obtained in MHD. However, slight differences are visible in the case of model 4, with the magnetic field corresponding to CGL-MHD being more homogeneous than the MHD one.

Model 5, corresponding to the subsonic and super-Alfv\'enic weak turbulence regime, is shown in the second row of Figure \ref{f1b2}. The strong firehose instabilities triggered in this regime deform the magnetic field 
lines as in model 2, producing a very granulated distribution of the magnetic 
field intensity that is very different from the stucture found in MHD. 

The third row of Figure \ref{f1b2} shows the results obtained in model 6, corresponding to the supersonic and sub-Alfv\'enic strong turbulence regime. In this case, the evolution of turbulence causes a mirror instability in most of the computational domain. The
instability, although it is somewhat suppressed by strong turbulent motion, is responsible here for slowing the gas and reducing the effective sonic Mach number. Note, however, that the differences between the structure of the magnetic field intensity in MHD and CGL-MHD are not large, being the CGL-MHD case slightly more homogeneous than its MHD counterpart.

As it can be seen from Figures \ref{f1b1} and \ref{f1b2}, the late-time configuration of the magnetic field intensity is appreciably different not only among the different models, but also between MHD and CGL-MHD within each model. In particular, the differences in the magnetic field structure obtained in both formalisms is most significant in models 1, 2 and 5. This is a direct consequence of the fact that in these models the turbulence is weak, so the evolution of the instabilities is not strongly affected by turbulent motions. Moreover, as shown in Ref. \citet{Kowal11}, since these models are also subsonic the density and velocity spectra computed from CGL-MHD simulations show enhanced power at small scales as compared to the ones obtained in MHD. 

In the next section, we shall relate the features discussed above to the values of the diffusion coefficients corresponding to a particle transversing the plasma.

\subsection{Diffusion coefficients}

\begin{figure}
\begin{center}
\includegraphics*[width=9.0cm]{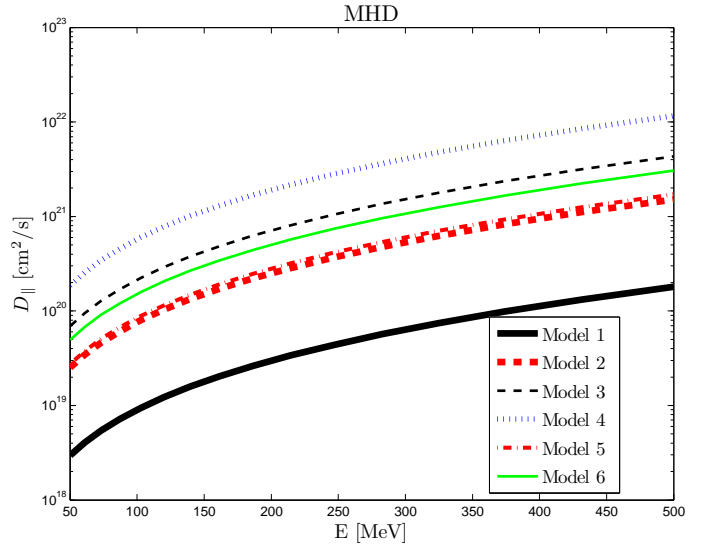}
\end{center}
\caption{$D_\parallel$ as a function of proton energy, as obtained from MHD numerical simulations for the six models considered.}
\label{figure7}
\end{figure}

\begin{figure}
\begin{center}
\includegraphics*[width=9.0cm]{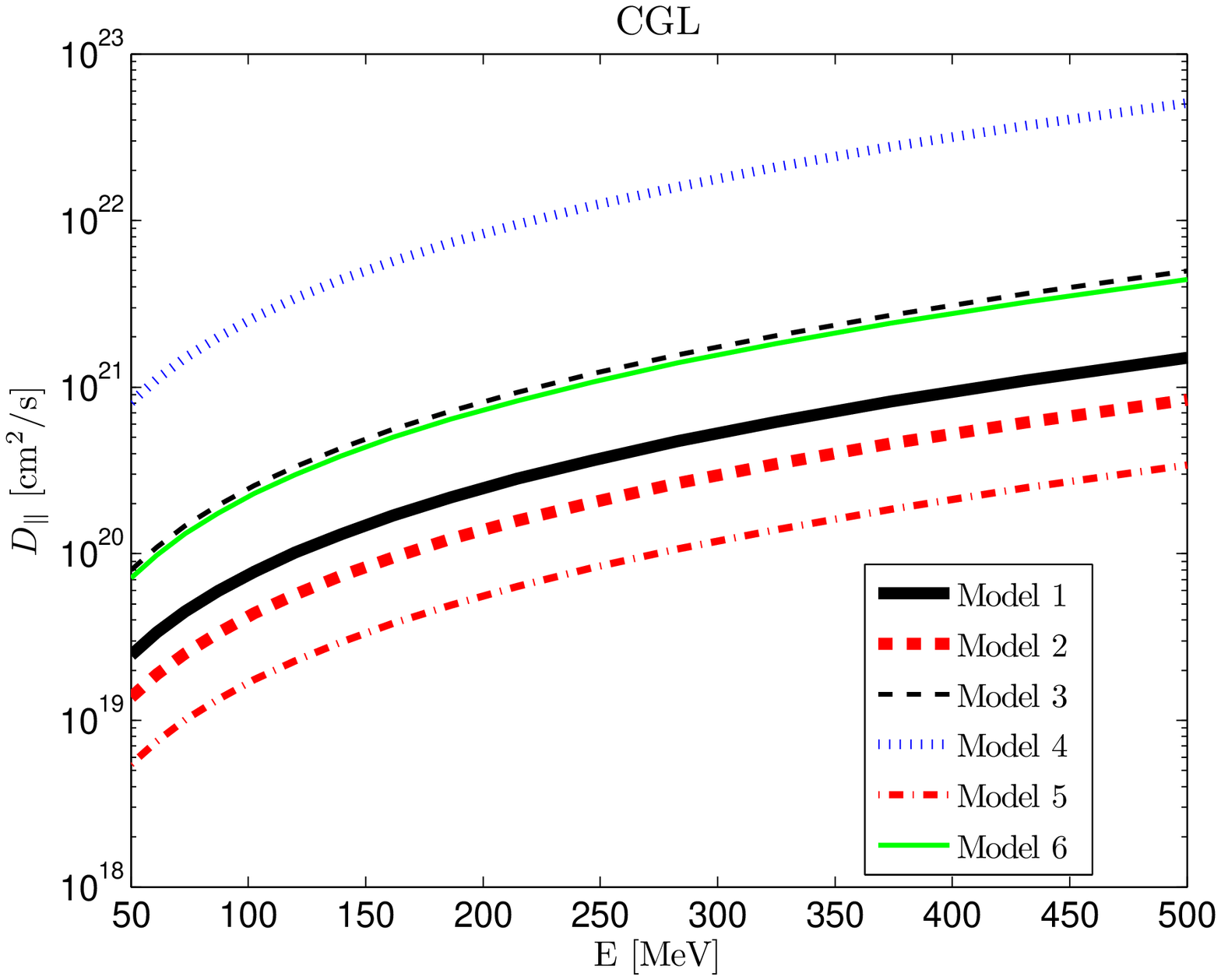}
\end{center}
\caption{$D_\parallel$ as a function of proton energy, as obtained from CGL-MHD numerical simulations for the six models considered.}
\label{figure8}
\end{figure}

\begin{figure}
\begin{center}
\includegraphics*[width=9.0cm]{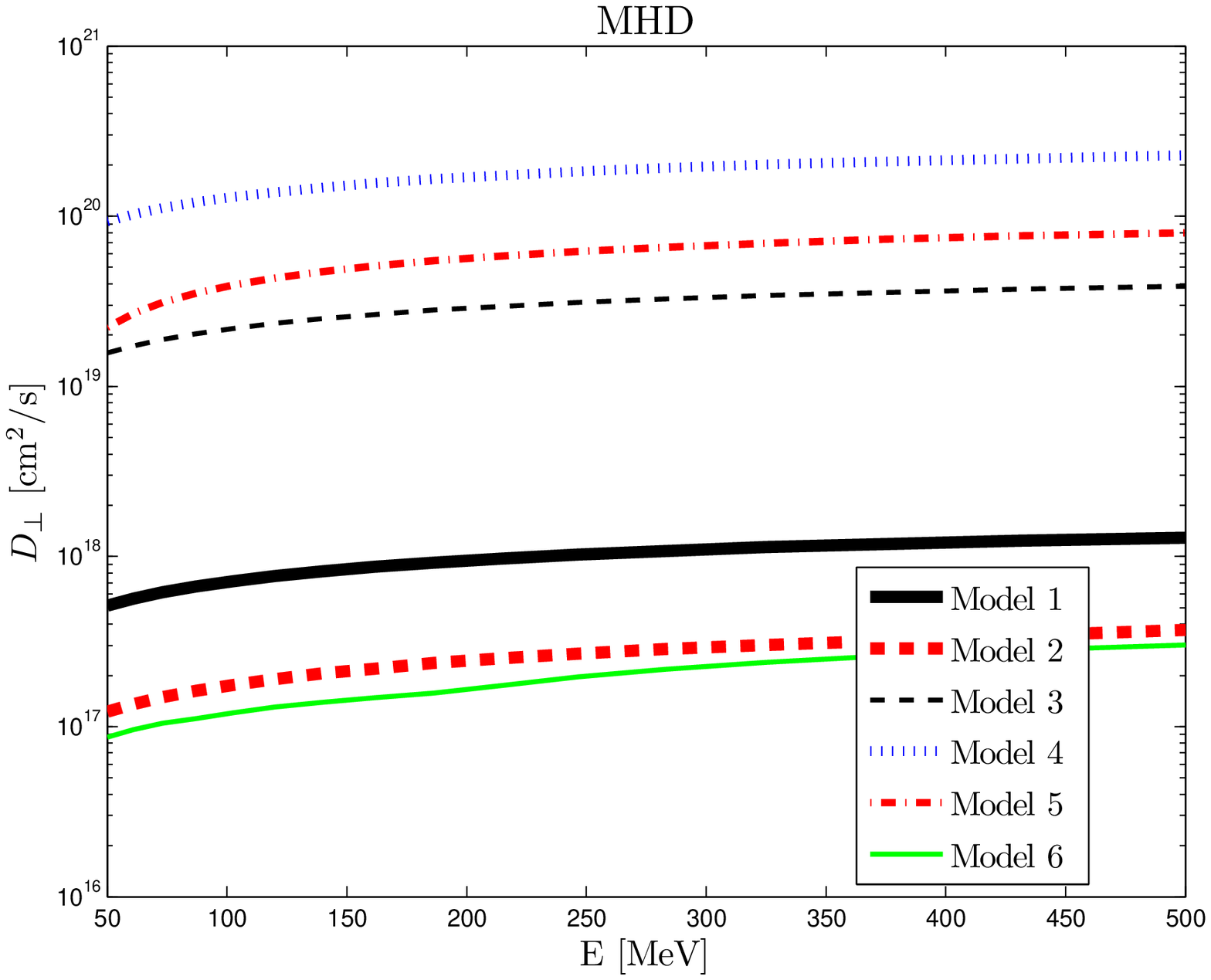}
\end{center}
\caption{$D_\perp$ as a function of proton energy, as obtained from MHD numerical simulations for the six models considered.}
\label{figure9}
\end{figure}

\begin{figure}
\begin{center}
\includegraphics*[width=9.0cm]{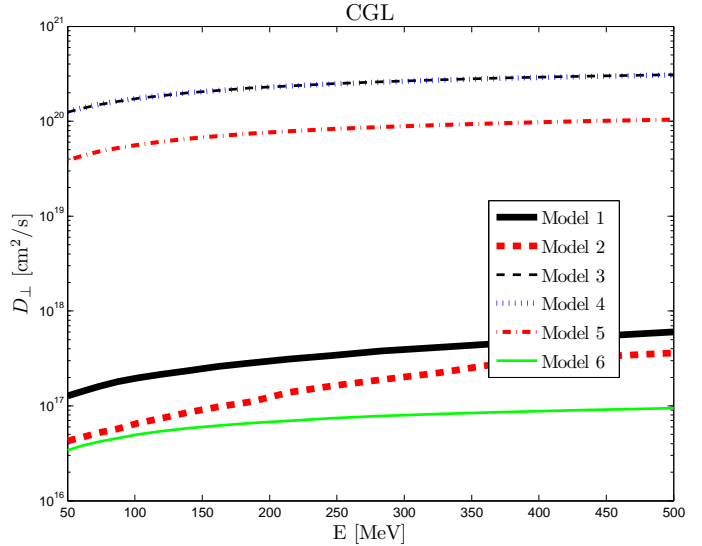}
\end{center}
\caption{$D_\perp$ as a function of proton energy, as obtained from CGL-MHD numerical simulations for the six models considered.}
\label{figure10}
\end{figure}

We start by comparing the diffusion coefficients obtained in the different models considered, in MHD and in CGL-MHD. Figures \ref{figure7} and \ref{figure9} show $D_\parallel$ and $D_\perp$ as a function of proton energy, as obtained from MHD simulations. Figures \ref{figure8} and \ref{figure10} show the same quantities but corresponding to CGL-MHD simulations. 

The first thing to notice from Figures \ref{figure7}-\ref{figure10} is that the values of $D_\parallel$ and $D_\perp$ are appreciably sensitive to the plasma parameters that determine the turbulent regime. The differences between the values of $D_\parallel$ and $D_\perp$ corresponding to the six models considered can be as large as three orders of magnitude. This shows that the diffusion coefficients computed from first-order QLT depend significantly on the turbulent regime in which the collisionless plasma evolves. 

It is worth noting also that the dependence of the diffusion coefficients on energy is practically independent of the plasma regime. This result is in line with that of Ref. \citet{Shalchi09}, where it was shown that a non-Gaussian magnetic field distribution only affects the amplitude of diffusion coefficients.  
From Figures \ref{figure7}-\ref{figure10}, it is seen that $D_\parallel$ increases with increasing energy, while $D_\perp$ is weakly dependent on energy. The values that we obtain for $D_\parallel$ and $D_\perp$ are of the order of those reported in previous observational and numerical studies (see, e.g., \citet{Giacalone98,JoeBOOK}), but we emphasize that it is not our purpose here to make detailed comparisons to other cosmic-ray transport models.

As it can be seen from Figure \ref{figure7}, in MHD the parallel diffusion is largest in model 4, corresponding to the supersonic and super-Alfv\'enic strong turbulence regime, and least important in model 1, corresponding to the subsonic and sub-Alfv\'enic weak turbulence regime. 
In CGL-MHD (Figure \ref{figure8}), parallel diffusion is again larger in model 4, but smallest in model 5 instead, that corresponds to the subsonic and super-Alfv\'enic weak turbulence regime with firehose instabilities. The latter feature may be due to the fact that in model 5, firehose instabilities can grow freely and tend to produce more granulated structures along the direction of the mean magnetic field, so that the parallel diffusion suffered by a particle moving parallel to the mean magnetic field is, on average, reduced.
 
For $D_\perp$ in MHD (shown in Figure \ref{figure9}), the largest values are obtained once again in model 4, while the smallest ones are obtained in model 6, corresponding to the supersonic and sub-Alfv\'enic strong turbulence regime. In CGL-MHD (Figure \ref{figure10}), perpendicular diffusion is most important in model 3 (supersonic and super-Alfv\'enic strong turbulence regime) and model 4, being the values in these two models very close to each other. The smallest values of $D_\perp$ are again obtained in model 6. The fact that $D_\perp$ is smallest for model 6 both in MHD and CGL-MHD can be explained by noting that in this model the local magnetic field grows as turbulence develops, resulting in strong magnetic breaking. In contrast to what happens in model 5, in model 6 the magnetic field is strong so that the perpendicular flows are not capable of destroying the magnetic configuration generated by strong turbulence. We stress that this behavior is not related to kinetic instabilities, but rather to the strong turbulence regime.

We shall now focus on the differences between the results obtained from MHD and from CGL-MHD. From Figures \ref{figure7} and \ref{figure8}, it is seen that the values of $D_\parallel$ corresponding to models 1 and 4 in CGL-MHD are  enhanced by an order of magnitude as compared to MHD. In contrast, the values corresponding to models 2 and 5 in CGL-MHD are somewhat suppressed, probably due to the effect of firehose instabilities which in these two models can grow freely and generate small-scale fluctuations of the magnetic field along the direction of the mean magnetic field. The values of $D_\parallel$ corresponding to models 3 and 6 do not change much when going from MHD to CGL-MHD.

Going over to $D_\perp$, we see from Figures \ref{figure9} and \ref{figure10} that the values of $D_\perp$ corresponding to models 1, 2 and 6 are somewhat suppressed in CGL-MHD as compared to MHD. In contrast, the value of $D_\perp$ corresponding to model 3 is enhanced by one order of magnitude in CGL-MHD as compared to MHD. 
It is seen also that the values of $D_\perp$ corresponding to models 4 and 5 are roughly the same in both formalisms.

\begin{figure}
\begin{center}
\includegraphics*[width=9.0cm]{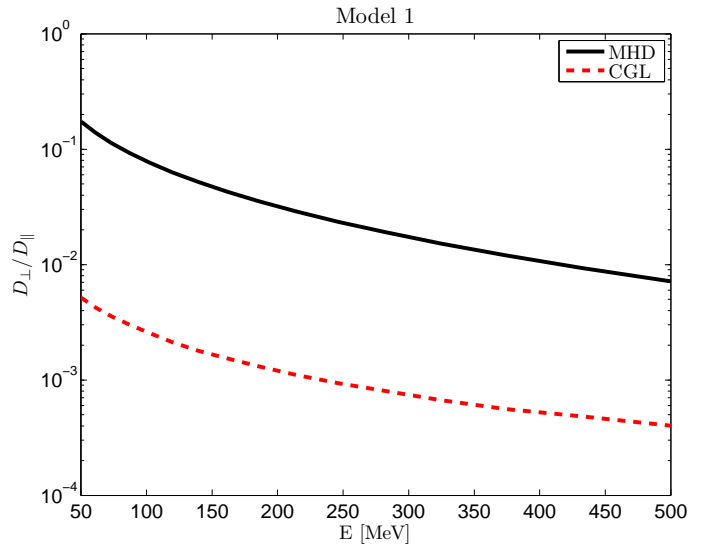}
\end{center}
\caption{$D_\perp/D_\parallel$ as a function of proton energy, as obtained from model 1 in MHD and CGL-MHD.}
\label{figure1}
\end{figure}

\begin{figure}
\begin{center}
\includegraphics*[width=9.0cm]{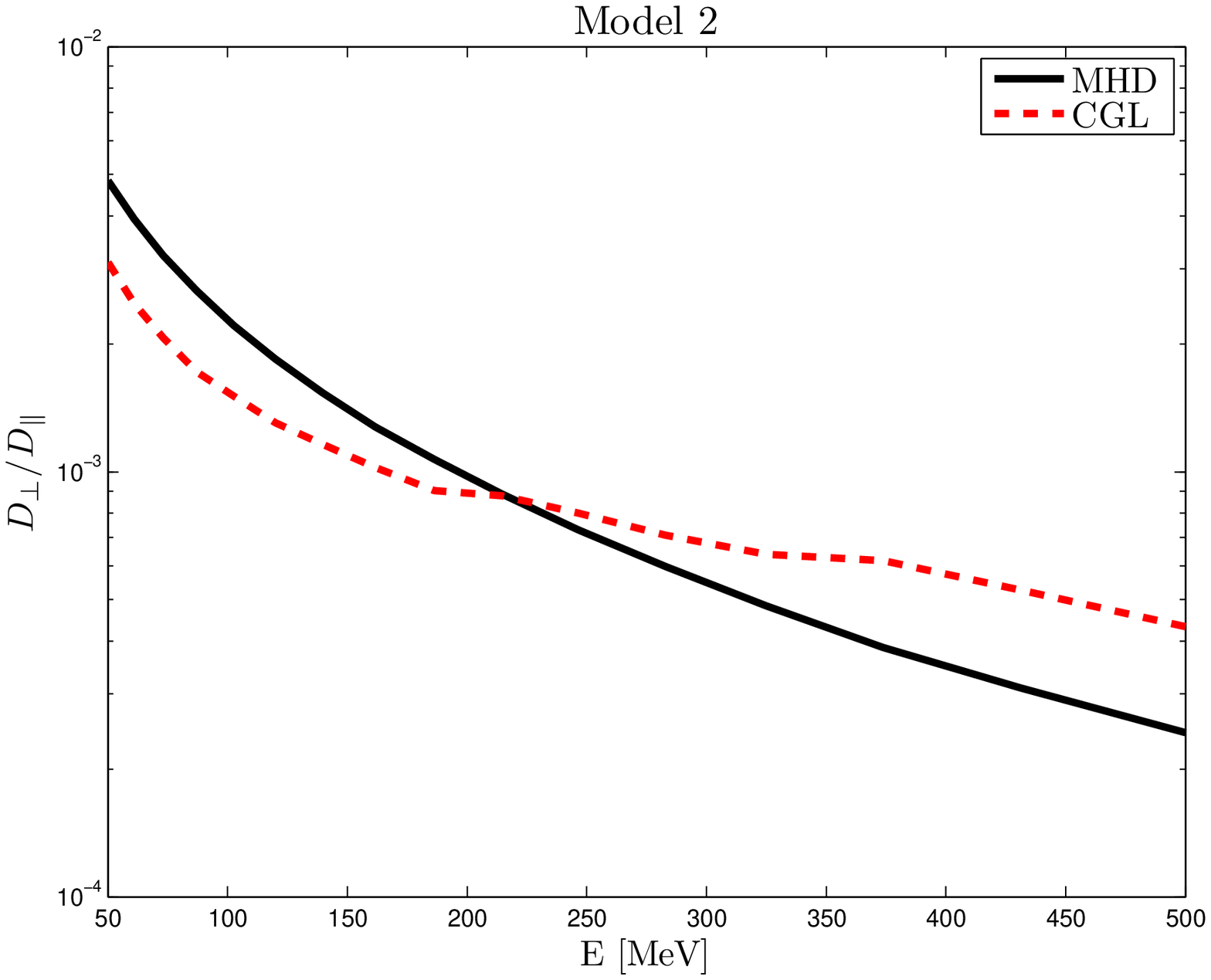}
\end{center}
\caption{$D_\perp/D_\parallel$ as a function of proton energy, as obtained from model 2 in MHD and CGL-MHD.}
\label{figure2}
\end{figure}

\begin{figure}
\begin{center}
\includegraphics*[width=9.0cm]{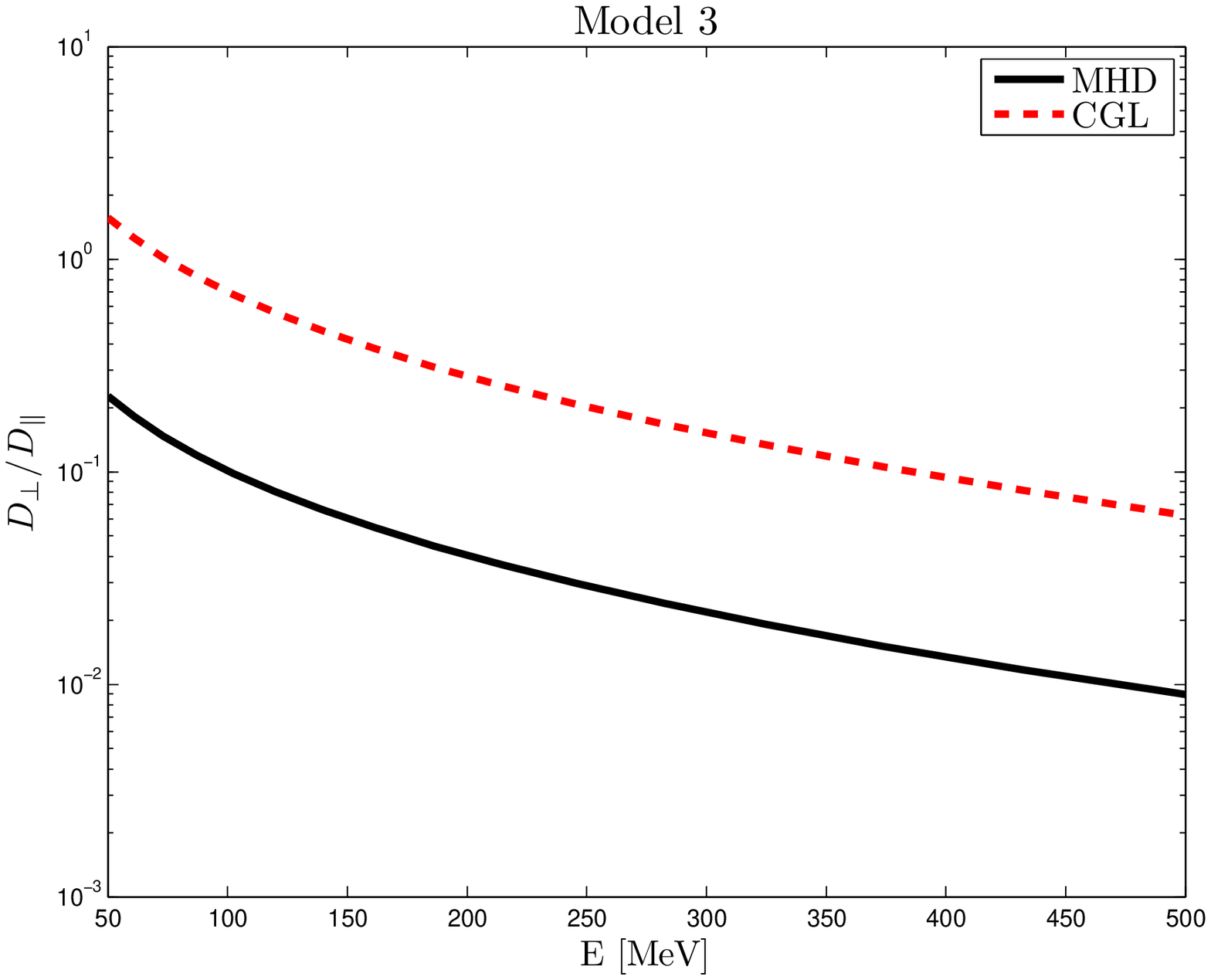}
\end{center}
\caption{$D_\perp/D_\parallel$ as a function of proton energy, as obtained from model 3 in MHD and CGL-MHD.}
\label{figure3}
\end{figure}

\begin{figure}
\begin{center}
\includegraphics*[width=9.0cm]{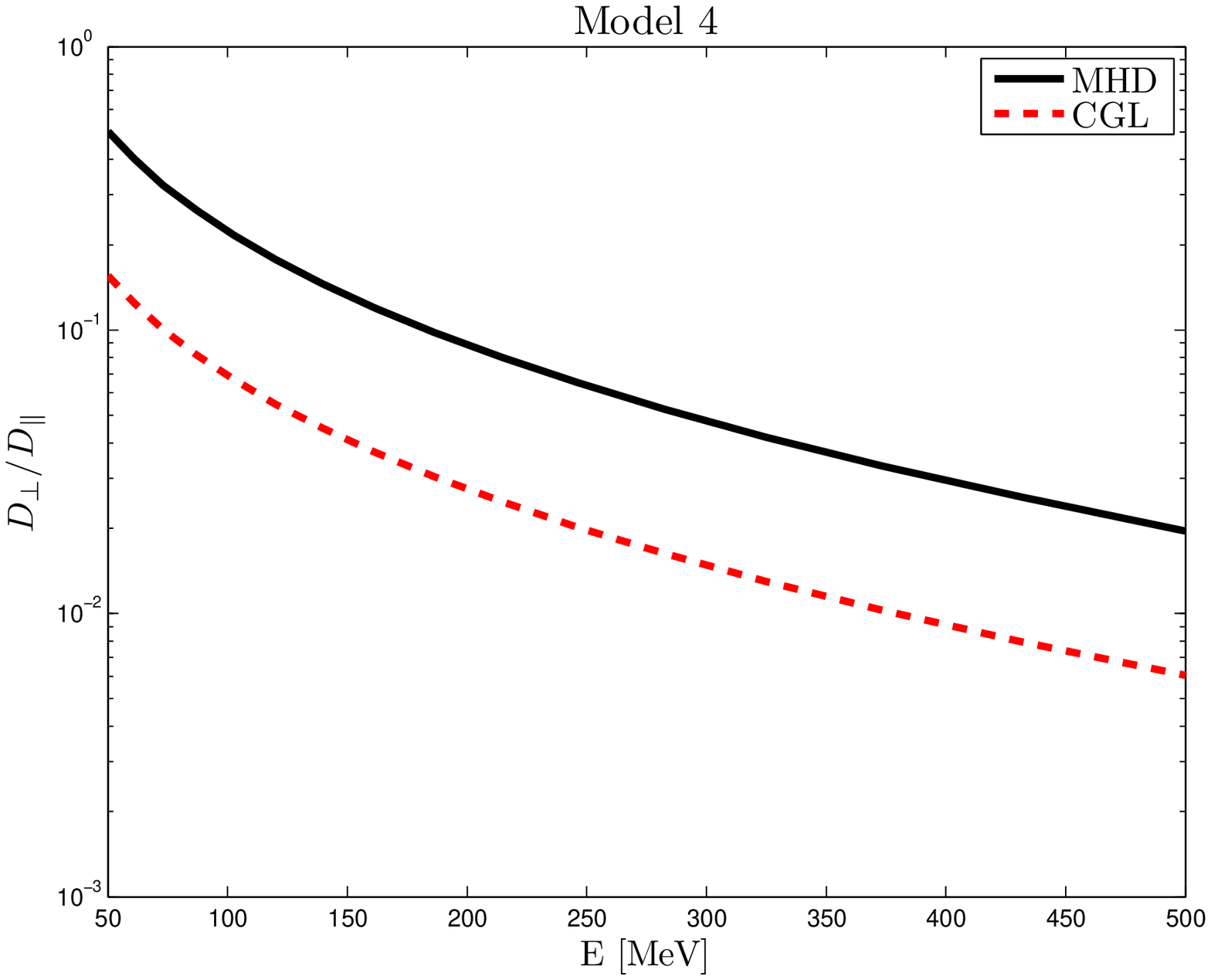}
\end{center}
\caption{$D_\perp/D_\parallel$ as a function of proton energy, as obtained from model 4 in MHD and CGL-MHD.}
\label{figure4}
\end{figure}

\begin{figure}
\begin{center}
\includegraphics*[width=9.0cm]{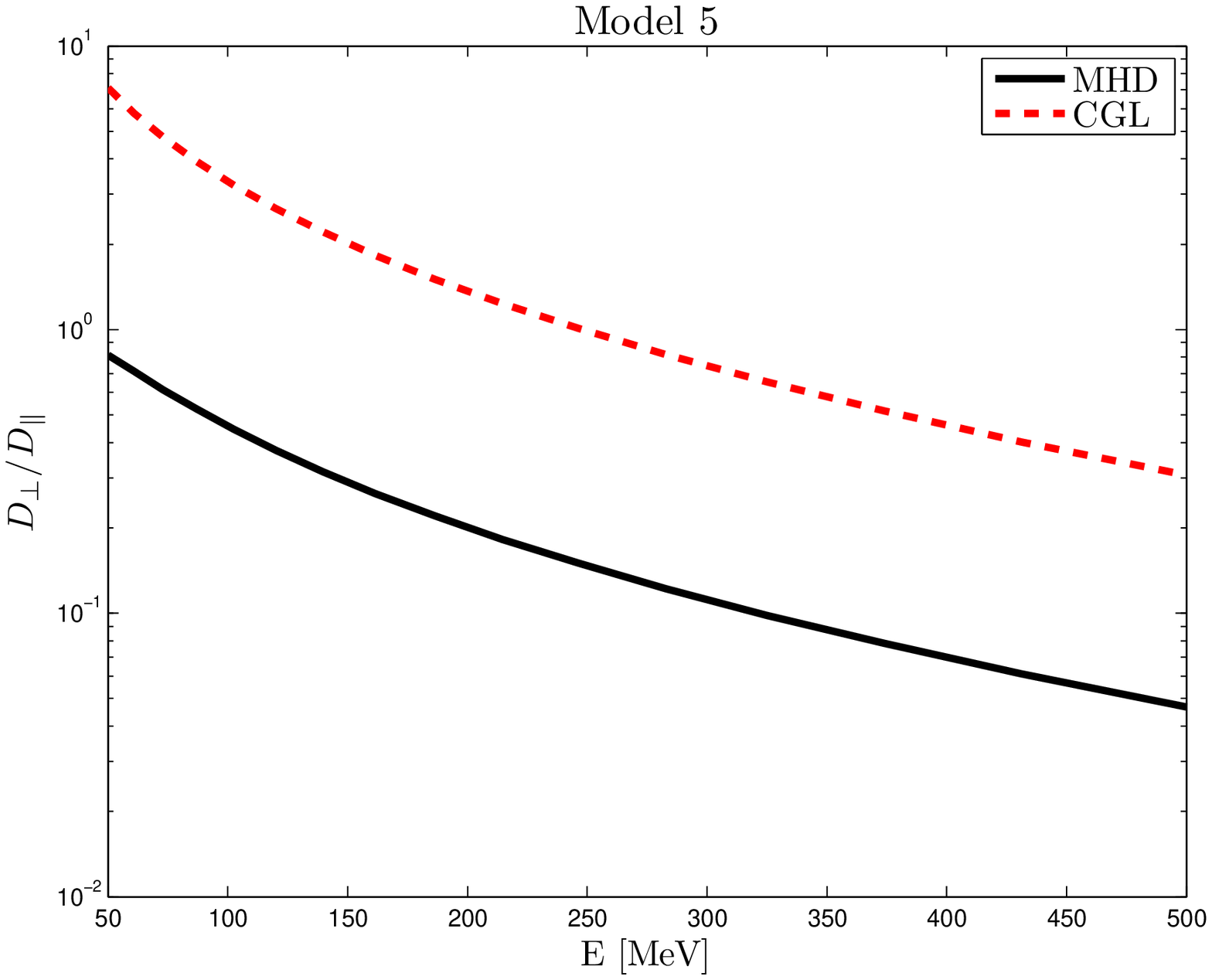}
\end{center}
\caption{$D_\perp/D_\parallel$ as a function of proton energy, as obtained from model 5 in MHD and CGL-MHD.}
\label{figure5}
\end{figure}

\begin{figure}
\begin{center}
\includegraphics*[width=9.0cm]{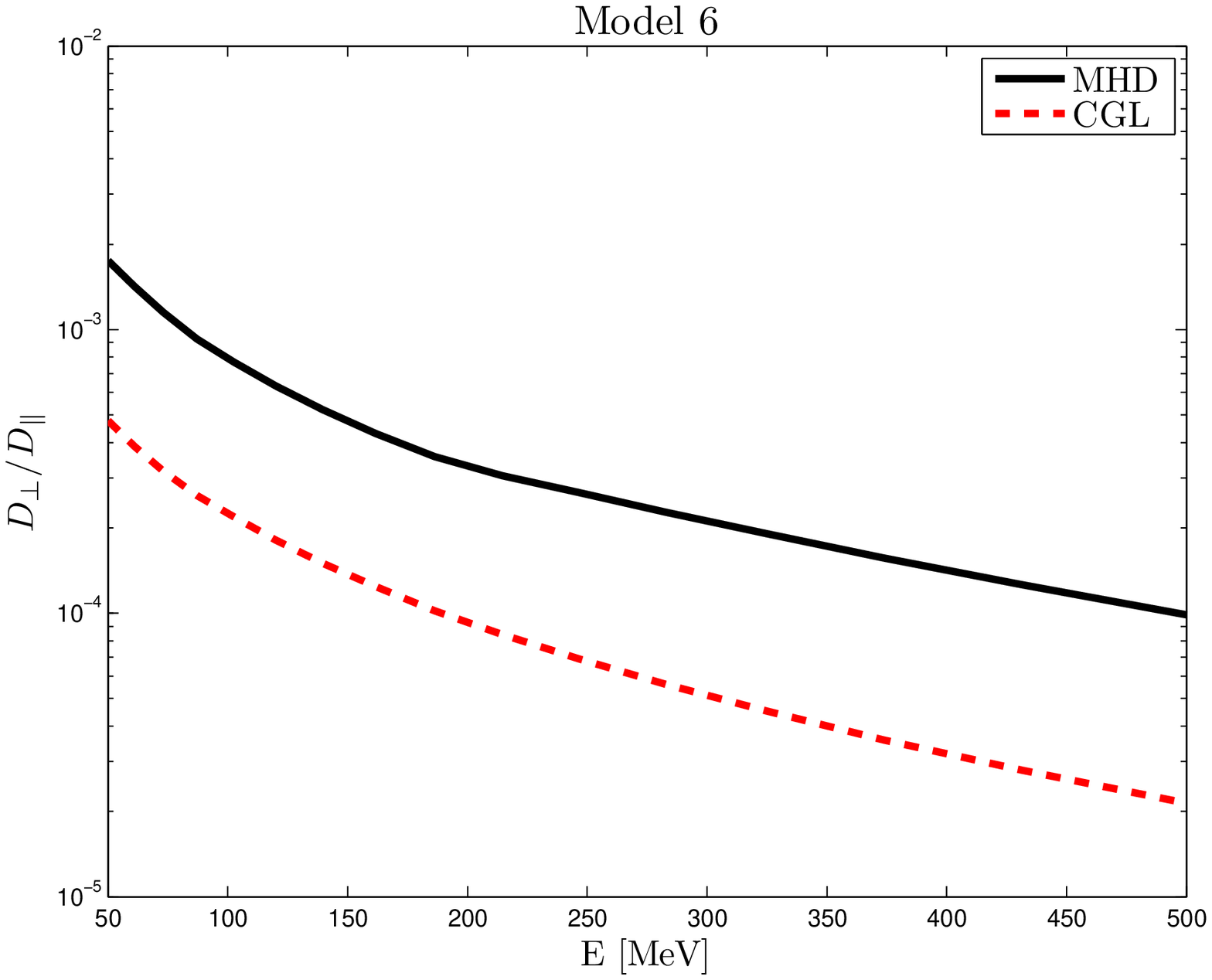}
\end{center}
\caption{$D_\perp/D_\parallel$ as a function of proton energy, as obtained from model 6 in MHD and CGL-MHD.}
\label{figure6}
\end{figure}

Some insight into the effect of pressure anisotropy on the diffusion coefficients can be achieved by comparing the ratio of perpendicular to parallel diffusion coefficients, $D_\perp/D_\parallel$, that is obtained in the two formalisms. Figures \ref{figure1}-\ref{figure6} show $D_\perp/D_\parallel$ as a function of proton energy, as obtained from each of the six models for MHD and CGL-MHD. 

It is seen that the ratio decreases with increasing particle's energy. $D_\perp/D_\parallel$ typically decreases by approximately one order of magnitude in the energy range considered. 

For models 1, 4 and 6, shown in Figures \ref{figure1}, \ref{figure4} and \ref{figure6} the ratio in MHD larger than the one in CGL-MHD. Note that in models 1 and 4 the ratio in MHD is larger than $10^{-2}$, while in model 6 the ratio is smaller than $\sim 10^{-3}$.
In contrast, for models 3 and 5 (Figures \ref{figure3} and \ref{figure5}) the ratio in CGL-MHD is larger than the one in MHD. In model 5, the ratio in CGL-MHD is close to $\sim 10^1$ at the lowest energies considered, while the ratio in MHD is roughly the same as in model 4. For model 3, the ratio reaches $\sim 10^0$ in CGL-MHD and $\sim 10^{-1}$ in MHD. 

For model 2, shown in Figure \ref{figure2}, the ratios corresponding to MHD and CGL-MHD are very similar to each other and rather small $\sim 10^{-3}-10^{-2}$, and become almost indistinguishable to the accuracy of our calculations. The similarity of the values of $D_\perp/D_\parallel$ in MHD and CGL-MHD that we obtain in model 2 is due to the effect of the firehose instabilities on the magnetic field fluctuations. Model 2 corresponds to a weak turbulence regime in which these instabilities can grow freely, and whose main effect is to isotropize the magnetic field fluctuations. The isotropization of the magnetic field fluctuations is not evident in models 4 and 5, which also involve firehose instabilities. In model 4 this is because the turbulence is strong enough to destroy the configuration attained by the effect of the firehose instabilities. In model 5, the magnetic field is weak (whereas in model 2 it is strong), and therefore the motion perpendicular to the local magnetic field lines is able to destroy the magnetic field configuration generated by the firehose instabilities. 

The largest difference in the values of the ratio obtained in MHD and CGL-MHD is obtained in model 1, where it can reach almost two orders of magnitude. We recall that this model corresponds to a weak turbulence regime with mirror instabilities. These kinetic instabilities tend to increase the anisotropy of magnetic field fluctuations, thus leading to a larger difference between MHD and CGL-MHD. Model 3 also presents mirror instabilities, but the model corresponds to a strong turbulence regime, so that the effect of mirror instabilities is largely washed out by the turbulent motion of the plasma.

\section{Conclusions}
\label{con}

Within the first-order quasi-linear theory of particle propagation through a fluctuating magnetized plasma, we have computed the parallel and perpendicular diffusion coefficients for low energy protons travelling through a collisionless plasma. The turbulent magnetic field evolution was modeled using a hybrid kinetic-magnetohydrodynamics formalism that incorporates the effect of pressure anisotropy and its associated instabilities, or otherwise by standard magnetohydrodynamics. We have analysed six plasma models with parameters corresponding to the sub/supersonic and sub/super-Alfv\'enic regimes, leading to weak turbulence dominated by mirror or firehose instabilities, or else strong turbulence in which the magnetic configuration generated by instabilities is largely washed out.  

Our main result is that the diffusion coefficients calculated using magnetic fields obtained from MHD-like numerical simulations that include or not pressure anisotropy can differ substancially. Moreover, we have shown that the values of the diffusion coefficients depend significanlty on the regime in which the collisionless plasma evolves. 
These results prompts us to perform detailed studies of energetic particle propagation through collisionless turbulent space plasmas where pressure anisotropy is present, using for this purpose more accurate formalisms that remove the limitations of the first-order quasi-linear theory used here. Work is in progress in this direction.

We thank G. Kowal for making the numerical code {\it Godunov} available to us. 
This work has been supported in part by the Consejo Nacional de Investigaciones Cient\'ificas y T\'ecnicas (CONICET) of Argentina. 

\bibliographystyle{aa} 
\bibliography{biblioJ} 
 {\typeout{}
  \typeout{****************************************************}
  \typeout{****************************************************}
  \typeout{** Please run "bibtex \jobname" to optain}
  \typeout{** the bibliography and then re-run LaTeX}
  \typeout{** twice to fix the references!}
  \typeout{****************************************************}
  \typeout{****************************************************}
  \typeout{}}







\end{document}